\documentstyle[11pt,psfig]{article}

\begin{document}
\begin{titlepage}

\begin{center}
{\Large \bf 
Truncation Effects in \\
Monte Carlo Renormalization Group Improved Lattice Actions}

\vspace{1cm}
{Tetsuya Takaishi\footnotemark[1]
{\it and} Philippe de Forcrand\footnotemark[2]}

{
\small \it \footnotemark[1]Hiroshima University of Economics \\
\small \it Hiroshima 731-01 \\
\small \it  Japan \\

\small \it \footnotemark[2]Swiss Center for Scientific Computing  \\
\small \it  ETH-Z\"urich, CH-8092 Z\"urich \\
\small \it  Switzerland \\
}
\end{center}
\vspace{2cm}

\abstract{
We study truncation effects in the SU(3) gauge actions  
obtained by the Monte Carlo renormalization group method.
By measuring the heavy quark potential
we find that the truncation effects in the actions coarsen the lattice by 40-50\% from the original blocked lattice.
On the other hand, we find that rotational symmetry of the heavy quark 
potentials is well recovered on such coarse lattices, 
which may indicate that rotational symmetry breaking terms
are easily cancelled out by adding a short distance operator.
We also discuss the possibility of reducing truncation effects.
}
\end{titlepage}

\section{Introduction}
Recently a lot of attention has been devoted to improvements of lattice discretized actions.
There exist two approaches to improving actions. One is the perturbative improvement program suggested by 
Symanzik\cite{SYM} and the other the renormalization group improvement program by Wilson\cite{WILSON}.
Early attempts at the perturbative improvement program did not appear practical for Monte Carlo simulations\cite{Forcrand}.
Recently, however, it has been revitalized with the help of tadpole improvement\cite{TAD} and actively been 
investigated by Monte Carlo simulations.

The renormalization group improvement program is very attractive since it can, 
in principle, give us lattice-artifact-free actions. 
In practice, however, it is not an easy task to obtain such actions
since we do not know a practical  way to determine them.
Recent successful attempts\cite{PERFECT} approximate the perfect action, which is defined on the renormalized trajectory,
with the fixed point action ( classically perfect action ) obtained in the limit of $\beta \rightarrow \infty$. 
An advantage of this method is that the fixed point action is rather easily  obtainable. 
Although the fixed point action is an approximation to the perfect action, 
this approximation turns out to be rather good in Monte Carlo tests.

Direct attempts\cite{ATT,TAKAISHI,DEMON,DEMONtaka} 
to obtain actions on the renormalized trajectory have been also made, using the Monte Carlo renormalization group (MCRG) method\cite{MCRG}. 
The MCRG method is useful not only for determining improved actions 
but also for other purposes, like monitoring the flow of couplings under
a scale transformation.  
In lattice QCD for instance, the MCRG method has been successfully used 
for determining the coupling shift $\Delta\beta$ of the $\beta$ function 
to reveal the scaling behavior of the theory\cite{BETA}.

The MCRG approach uses the fact that the blocked trajectory generated by 
the blocking transformation reaches the renormalized trajectory 
after sufficiently many  blocking steps
and then runs along with the renormalized trajectory.
Therefore the configurations generated by successive blocking transformations
will correspond to an action located nearer and nearer the renormalized trajectory.
Such configurations should have less artifacts than 
non-blocked configurations since from Wilson's renormalization group argument
we expect that continuum physics is realized on the renormalized trajectory. 
Actually it has been  shown that rotational symmetry of 
the heavy quark potential is well recovered 
on blocked configurations\cite{TAKAISHI}. 
If we can determine the action representing the blocked configurations, 
we can directly generate the configurations without blocking.
The determination of the action was tackled by the canonical demon method\cite{DEMON,DEMONtaka} which produces as output a set of coupling constants.
In the canonical demon method, like other determination methods,
the action to be determined must be truncated to a certain local form, which may cause some error. 
We call the effects of this error {\it truncation effects}.
Unless the truncated form of the action is 
sufficiently close to the real one, 
truncation effects may affect long-distance, physical properties of the action. 
In Ref\cite{TAKAISHI} several SU(3) gauge actions corresponding to blocked configurations 
(generated by a blocking transformation)
 were obtained in multi-dimensional coupling space.
However it is not known whether these actions still preserve the improvements of the blocked configurations or not,
since these improvements may have been ruined by truncation effects. 
This point can only be examined by Monte Carlo simulations.
In this letter, we perform Monte Carlo simulations with the actions 
obtained in Ref\cite{TAKAISHI}, 
which we call MCRG improved actions, 
examine the rotational symmetry of the heavy quark potential,
and estimate the truncation effects which appear in a physical quantity ( the string tension ).

\section{Monte Carlo Renormalization Group Improved Actions}
We briefly sketch the method used to determine the MCRG actions in Ref\cite{TAKAISHI}.
First, we block configurations generated with the standard Wilson action on $32^3\times 64$ lattices.
The blocking scheme employed is Swendsen's scale factor 2 blocking scheme.
This blocking scheme was optimized by multiplication by a Gaussian random SU(3) matrix so that 
the blocked trajectory converges to the renormalized trajectory quickly\cite{QCDTARO,TAKAISHI}.
Actually the optimal width of the Gaussian distribution at $\beta \sim 6.0$ turned out to be 0, 
meaning that the Gaussian random SU(3) matrix becomes unity. This optimal width was used in the blocking.
The blocking was performed twice, starting from $32^3\times64$ lattices at
two $\beta$ values, 6.0 and 6.3,
thus resulting in two sets of $8^3\times 16$ blocked configurations. 

Next, in order to determine coupling constants 
we apply the canonical demon method\cite{DEMON,DEMONtaka} on the blocked configurations.
The canonical demon method introduces several degrees of freedom, so-called demons, which are associated with 
each of the coupling constants to be determined.
The action $S$ can be written as
\begin{equation}
S=\sum_i \beta_i \bar{S}_i(U)
\label{demoneq0}
\end{equation}
where $\beta_i$ is a coupling constant and $\bar{S}_i(U)$ is some operator consisting of Wilson loops.
Hereafter for simplicity, 
let us assume that the action is characterized by only one operator or 
by one coupling constant:
\begin{equation}
S= \beta \bar{S}(U).
\end{equation}
The demon is updated by a microcanonical simulation in the joint system, 
i.e. the demon and the links of a blocked configuration.
In the microcanonical simulation, the total energy of the joint system,
i.e. $\bar{S}(U)$ plus the demon energy $E_d$, is kept constant.

In the canonical demon method,
in order to avoid a possible finite volume error,
a set of well uncorrelated blocked configurations is prepared.
At a certain stage of the microcanonical simulation, 
we move to a new blocked configuration chosen from 
the set and the configuration used before is discarded.
The demon is also moved to the new configuration 
keeping the value of the demon energy at the last update.  

The probability distribution of the demon energy $P(E_d)$ in the simulation 
is expected to be the Boltzmann distribution:
\begin{equation}
P(E_d) \sim \exp (-\beta E_d)
\label{demoneq}
\end{equation}
where  $\beta$ is the coupling constant to be determined.
Using eq.(\ref{demoneq}), we write
\begin{equation}
<E_d>= \int_{E_{min}}^{E_{max}} E_d \exp(-\beta E_d) dE_d/Z
\label{demoneq2}
\end{equation}
where the demon energy $E_d$ is restricted in the region, 
$E_{min} < E_d < E_{max}$
and $Z$ is the partition function.
If we take $E_{max}=-E_{min}=E_c$, where $E_c$ stands for some constant value
which we fix in the simulation, eq.(\ref{demoneq2}) will be
\begin{equation}
<E_d>= 1/\beta - E_c/tanh(\beta E_c).
\label{demoneq3}
\end{equation}
Finally, substituting the average value $<E_d>$ obtained from the simulation,
we determine the value of the coupling constant $\beta$ by solving eq.(\ref{demoneq3}) numerically.
The extension to the multi-coupling form of eq.(\ref{demoneq0}) is straightforward.

In our study we take MCRG actions obtained in two-dimensional coupling space.
The actions are written as
\begin{equation}
S=Re(\beta_{11} \sum Tr(1\times1\: 
{\sf Wilson \:loop})/3 + \beta_{12} \sum Tr(1\times2\: {\sf Wilson \:loop})/3).
\end{equation}
The values of the couplings $\beta_{11}$ and $\beta_{12}$ are listed in Table 1.
The actions {\bf A} and {\bf B} come from Ref\cite{TAKAISHI}, 
which are obtained from the configurations blocked twice at $\beta=6.00$ and
$\beta=6.30$ respectively.
We also use an interpolated action ({\bf M}) located half-way between actions {\bf A} and {\bf B}.
Fig.1 shows the locations of these actions in the $\beta_{11}$-$\beta_{12}$ plane.
The ratio of the two couplings, $|\beta_{12}/\beta_{11}|$,
is bigger than that of the Symanzik tree-level improved actions\cite{WEISZ} and Iwasaki action\cite{IWASAKI}. 

\section{Heavy Quark Potentials}
We employ $8^3\times16$ lattices which is the same lattice size with the original blocked lattices. 
We calculate static potentials between a heavy quark and antiquark pair from the exponential fall-off of Wilson loops.
Our calculation is based on 500-700 configurations separated by 100-200 pseudo heat-bath sweeps.
The smearing technique is used to reduce errors in the extracted potentials.

By construction, the canonical demon method preserves the average value of the Wilson loops included in the action. 
This means that in our case the average values of $1 \times 1$ and $1\times2$ Wilson loops\footnote{
Some average values of Wilson loops of the MCRG actions are listed in Table 2.}, 
which are associated with 
$\beta_{11}$ and $\beta_{12}$, should be the same values as the blocked ones within statistical uncertainty.
Fig.2 and 3 show comparisons of $1\times T$ Wilson loops between the original blocked lattices and the lattices
obtained with the actions  {\bf A} and {\bf B}.
As seen in both figures, 
good agreement is observed for each of 
$1 \times 1$ and $1\times2$ Wilson loops. 
This check ensures that the canonical demon method worked correctly.
For other loops, however, the difference increases as $T$ increases.
 
In order to compare the potentials 
we plot  the potentials of the action {\bf A(B)} and original blocked lattice 
on the same figure(Fig.4-5).
It is clearly seen that the potentials of the action {\bf A(B)} are very different 
from the ones of the original blocked lattice, 
and the slopes of the potentials at large distance 
of the action {\bf A(B)} are larger,
which implies that the lattice spacings are also larger than the original ones.
We quantify this effect  by comparing string tensions.

The string tensions are extracted by fitting the potentials to the form
\begin{equation}
V_L(r)= m + \sigma_L r - \frac{c}{r},
\label{POT}
\end{equation}
where $m$, $\sigma_L$ and $c$ are fitting parameters, and $r=R/a$  is the distance measured in lattice units.
The fits are performed on all the data including  on and off axis potentials.
The fit results are summarized in Table 3.

In order to evaluate the truncation effect, 
let us compare the string tensions with the ones of the original blocked lattices.
The results are  the following:

\begin{equation}
\sigma_L = \left\{
\begin{array}{ll}
1.29(12) & \mbox{ MCRG Action \bf{A}} \\
0.58(4)  & \mbox{ Original Blocked Lattice(at $\beta=6.00$)}
\end{array}
\right.
\label{steq1}
\end{equation}

\begin{equation}
\sigma_L = \left\{
\begin{array}{ll}
0.587(27) & \mbox{ MCRG Action \bf{B}} \\
0.287(26)  & \mbox{ Original Blocked Lattice(at $\beta=6.30$)}.
\end{array}
\right.
\label{steq2}
\end{equation}

Here we comment on the naive expected string tension on a blocked lattice at $\beta=6.00$ and 6.30.
From the literature\cite{BALI}, we find $\sigma a^2\simeq 0.051$ at $\beta=6.00$ 
and $\sigma a^2\simeq 0.020$\footnote{We interpolate the value of the string tension for $\beta=6.30$.
since we do not find any Monte Carlo result at $\beta=6.30$.}
at $\beta=6.30$. Since we block twice, the naive expected string tension on the blocked lattice  will be 
$\sigma a^2\simeq 0.82(0.32)$ at $\beta=6.00(6.30)$ respectively. 
These values are compatible with that of  the blocked lattice at $\beta=6.30$ but not compatible at $\beta=6.00$
 ( See eqs.(\ref{steq1}) and (\ref{steq2}) ).
Probably this mismatch is due to the very small correlation length at $\beta=6.00$ 
since the correlation length $\xi$ is estimated to be $\xi\sim 4$ using $1/\xi^2 \simeq \sigma a^2$ and
this  correlation length is too small to preserve the same long range physics under the blocking transformation.

We examine the truncation effects which are seen in the string tension
by taking the ratio of the string tensions of the MCRG actions and the 
original blocked lattices (eqs. \ref{steq1},\ref{steq2}). 
\[
\frac{(\sigma_L)_{\bf MCRG}}{(\sigma_L)_{Blocked}} \approx \left\{
\begin{array}{ll}
2.22 & \mbox{  MCRG Action \bf{A}}  \\
2.04 & \mbox{  MCRG Action \bf{B}}.
\end{array}
\right.
\]

Since $\sigma_L=\sigma a^2$, the ratio in lattice spacing will be:
\begin{equation}
\frac{(a)_{\bf MCRG}}{(a)_{Blocked}} \approx  \left\{
\begin{array}{ll} 
1.5 & \mbox{ MCRG action \bf{A}}  \\
1.4 & \mbox{ MCRG action \bf{B}}.
\end{array}
\right.
\end{equation}

If the truncation effects were negligible, the ratio should be one.
However this is far from being the case, which indicates that sizeable truncation effects
are involved in the MCRG actions.
It is important to notice here is that the truncation effect {\em increases} the lattice spacing,
i.e. the lattice spacing of the MCRG actions is $40-50\%$ bigger than that of the original blocked lattice.
This increase can be understood in the following way. The MCRG actions contain only the coupling 
constants associated with small Wilson loops ($1\times1$ and $1\times2$ Wilson loops). 
On the other hand, the real action of the blocked configurations can have many coupling constants 
associated with large Wilson loops.
Let us assume that all the long range coupling constants are positive.
This assumption seems to be valid at least for coupling constants 
up to eight-links Wilson loop operators as found in \cite{TAKAISHI}, 
where all those coupling constants are positive.
The positive coupling constants constrain Wilson loops to be more ordered.
Since the large Wilson loops of the MCRG action are less ordered 
average values of the Wilson loops decay faster with distance than that of the constrained Wilson loops 
on the blocked configurations, 
as seen in Fig.2-3.
Thus the potential of the MCRG action, $V(R)_{MCRG}$, will be larger than that of the blocked configurations, 
$V(R)_{block}$:
\begin{equation}
V(R)_{MCRG} > V(R)_{block}.
\end{equation} 
This situation may become more pronounced at large $R$ as seen in Fig.4-5.
Thus, we obtain a larger string tension as extracted from the slope of the potential at large distance.

In order to examine the rotational symmetry of the potentials
we plot all the data of the MCRG action {\bf A},{\bf B} and {\bf M} on the same figure ( Fig.6 ) by rescaling them 
using the obtained string tension results.
The potentials are shifted so that  each fitted curve for  
$V(R)/\sqrt{\sigma}$ gives the value 2 at $R\sqrt{\sigma}=2$. 
We see  good rotational symmetry except at large $R\sqrt{\sigma}$ 
due to the large error bars.
The recovery of rotational symmetry will be more remarkable
if we compare the potentials with that of the standard Wilson action\cite{FIG}
where rotational symmetry is severely violated on a coarse lattice.

\section{Discussion}

The MCRG improvement approach has severe difficulties in determining 
the effective action.
When we determine the action, we have to truncate 
it to a certain local form.  
On the other hand the truncation effects may ruin some benefits 
from the MCRG improvement.
In order to reduce the truncation effects, 
one can tune the blocking transformation in such a way that 
contributions of long range coupling constants disappear quickly.
This tuning was successfully done for the fixed point action\cite{PERFECT}.
At finite $\beta$, however, a practical way to tune the blocking transformation is not known.
The brute force approach, i.e. to directly search for an optimal  blocking scheme 
by varying a couple of blocking parameters
in Monte Carlo simulations,
would certainly require a huge computational effort which we try to avoid.

Our MCRG improved actions have been obtained with a blocking scheme optimized by
a Gaussian random SU(3) matrix. 
Strictly speaking, however, this optimization does not guarantee that the truncation effect will be minimized.
Originally this optimization was introduced for the study of the $\Delta\beta$, and the optimization 
was done so that the blocked trajectory converges to the renormalized trajectory quickly. 
This quick convergence is only desirable for the matching method of the $\Delta\beta$ analysis. 
In this sense it is not sure that 
our MCRG actions are characterized by a local form. 
Actually our study found  a big difference in the lattice spacing by the analysis of the string tension. 
Even so good news were found: one of the evidences for improvement, 
recovery of the rotational symmetry, is observed even in the presence of the 
truncation effects.   

The one-loop level tadpole improvement scheme\cite{TAD} is known to work well, in which the action is designed to 
remove the ${\cal O}(a^2)$ and ${\cal O}(\alpha_s a^2)$ errors by adding two additional operators to the standard Wilson action. 
As far as rotational symmetry is concerned, the tree-level tadpole improvement scheme 
( Symanzik action + tadpole ) also works well,
in which one additional operator ($1\times2$ Wilson loop) is added to the standard Wilson action.
The only difference between the tree-level improvement ( Symanzik action ) with and without the tadpole scheme 
is that the tadpole scheme increases the negative coupling contribution of the $1\times2$ Wilson loop operator,
i.e., if we use the average plaquette value $<plaq>\simeq 0.4$, 
$\beta_{12}/\beta_{11}$ is approximately equal to $-0.08$, instead of 
$-0.05$ for the tree-level improvement scheme without tadpole improvement. 
For our MCRG actions, we find $\beta_{12}/\beta_{11}\simeq -0.1$.
The Iwasaki action, with which good rotational symmetry can be seen\cite{Iwasaki2}, also has 
similar behavior, i.e. $\beta_{12}/\beta_{11}\simeq -0.09$.
Therefore we suggest that rotational symmetry can be easily recovered with 
a short-distance correction ( $1\times2$ Wilson loop ) to the standard Wilson action
and the value of $\beta_{12}/\beta_{11}$  should be more negative 
than the naive tree-level value, i.e., $\beta_{12}/\beta_{11}= -0.08$ 
to $-0.1$ 
in a  region where the lattice spacing $a$ is as large as $a\simeq 0.3$
to $0.5$ fm.

The MCRG improved actions were determined with the canonical demon method.
One could try another scheme such as the Schwinger-Dyson (SD) equation method\cite{SD}. 
Truncation effects with the SD  equation method might be different, since in 
that case 
coupling constants are obtained by solving a set of equations 
whose coefficients are calculated from correlations among {\em arbitrary} Wilson loops, 
while the canonical demon method obtains coupling constants 
so that average values of Wilson loops included in the action {\em only} are conserved.
And even though the canonical demon method conserves the average values of such Wilson loops,
truncation effects already appear in the correlation among them\cite{DEMONtaka}.
Thus the SD equation might be more advantageous than the canonical demon method
since the SD equation uses additional information on the action
from the correlation among Wilson loops and determines the coupling constants by using  many available equations.
For the SU(3) gauge theory, determination of the renormalized trajectory was 
attempted with the SD equation method\cite{TARO}
and a rough picture of the RT was obtained.
It would be interesting to see how truncation effects appear in a different way 
for both the canonical demon and the Schwinger-Dyson equation methods. 

\vspace{1cm}
{\Large \bf Acknowledgments}
\vspace{.5cm}

We thank the QCD-TARO collaboration for their blocked configurations used in 
this study.
T.T. would like to thank the Swiss Center for Scientific Computing,
especially Professor Martin H. Gutknecht, for its hospitality.

\begin{table}
\begin{tabular}{ccc} \hline
 action    & $\beta_{11}$ & $\beta_{12}$ \\ \hline
  {\bf A}  & 6.1564      & -0.6241      \\ \hline
  {\bf M}  & 7.0712      & -0.7705     \\ \hline
  {\bf B}  & 7.986       & -0.9169     \\ \hline
\end{tabular}
\caption{
Values of coupling constants, $\beta_{11}$ and $\beta_{12}$, used in the study.
We label these actions by {\bf A}, {\bf M} and {\bf B} as in the table. }
\end{table}

\begin{table}
\footnotesize
\begin{tabular}{c|cc|cc|cc} \hline
Action  &       {\bf A}&              & {\bf M}      &              & {\bf B}      &                           \\ \hline         
$I\times J$  & 1 &  2           &  1           & 2            &   1          & 2                        \\ \hline 
1 & 0.409306(75) &              & 0.475636(77) &              & 0.538690(60) &                           \\ \hline
2 & 0.139237(62) & 0.010488(55) & 0.19660(10)  & 0.026427(54) & 0.263218(86) & 0.056835(92)              \\ \hline
3 & 0.045152(65) & 0.004230(39) & 0.071291(95) & 0.003129(62) & 0.127497(95) & 0.01255(10)   \\ \hline
4 & 0.014484(53) &              & 0.036781(82) & 0.000404(51) & 0.061726(93) & 0.002846(89)  \\ \hline  
5 & 0.004591(73) &              & 0.012743(91) &              & 0.029966(79) & 0.000674(77)              \\ \hline
6 & 0.001423(76) &              & 0.005091(71) &              & 0.014570(57) & 0.000136(77)              \\ \hline
7 & 0.000469(53) &              & 0.002000(91) &              & 0.007103(46) &                           \\ \hline
8 & 0.000155(43) &              & 0.000813(94) &              & 0.003505(69) &                           \\ \hline
\end{tabular}
\caption{
Average values of $I\times J$ Wilson loops of the actions {\bf A}, {\bf M} and {\bf B}. }
\end{table}

\begin{table}
\begin{tabular}{cccccc} \hline
  action   & m        &  $\sigma_L$ & c        & $\chi^2/d.o.f$ &  $a(fm)$  \\ \hline
  {\bf A}  & 0.11(29) & 1.29(12)    & 0.27(16) & 1.37         & 0.52(5) \\ \hline
  {\bf M}  & 0.35(22) & 0.885(85)   & 0.33(14) & 0.42         & 0.43(4) \\ \hline
  {\bf B}  & 0.39(7)  & 0.587(27)   & 0.251(40)& 1.17         & 0.35(2)  \\ \hline
\end{tabular}
\caption{Results of the fits. 
The lattice spacing $a$ is obtained by using $\sigma_L=\sigma a^2$ and $\surd\sigma=427MeV$.}
\end{table}

\newpage

\begin{figure}[htb]
\centerline{\psfig{figure=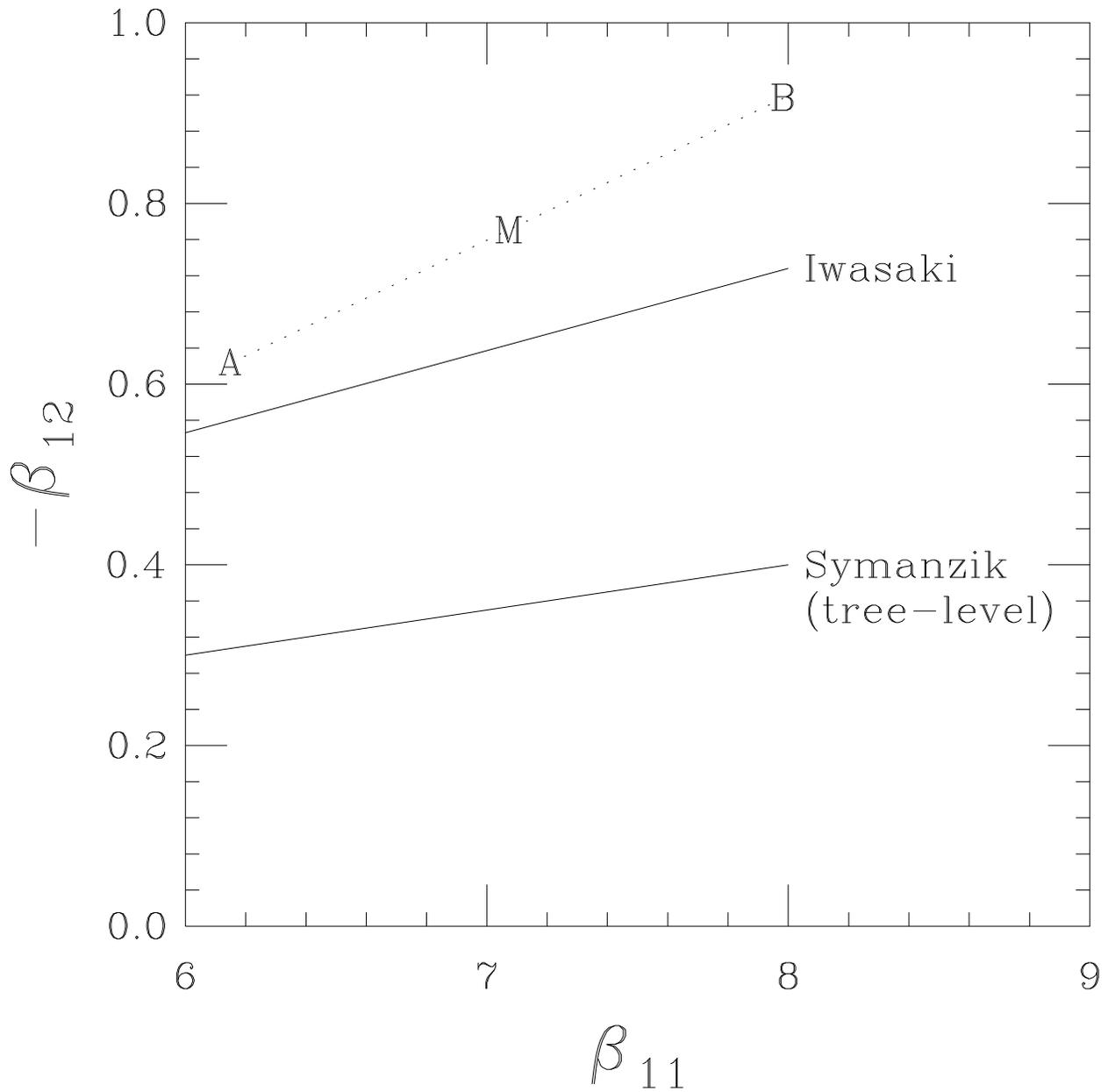}}
\caption{
Location of the actions in the $\beta_{11}-\beta_{12}$ plane.
Points {\bf A} and {\bf B} were obtained in [9].
Point {\bf M} is interpolated at the mid-point between {\bf A} and {\bf B}.
Symanzik and Iwasaki actions are also indicated by solid lines in the figure.
}
\end{figure}

\begin{figure}[htb]
\centerline{\psfig{figure=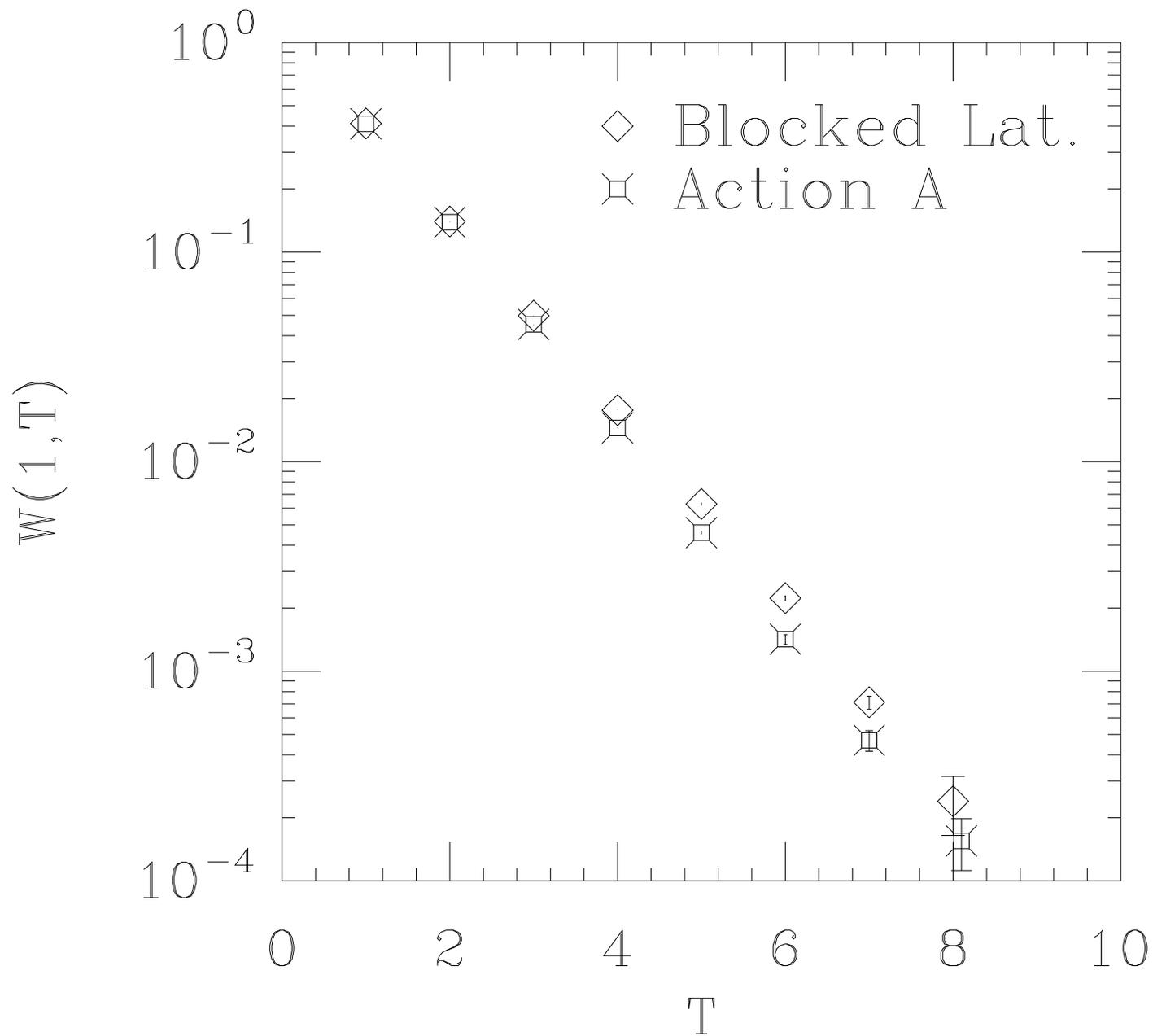}}
\caption{Comparison of $1 \times T$ Wilson loop values between action {\bf A} and 
the original blocked lattice. 
}
\end{figure}

\begin{figure}[htb]
\centerline{\psfig{figure=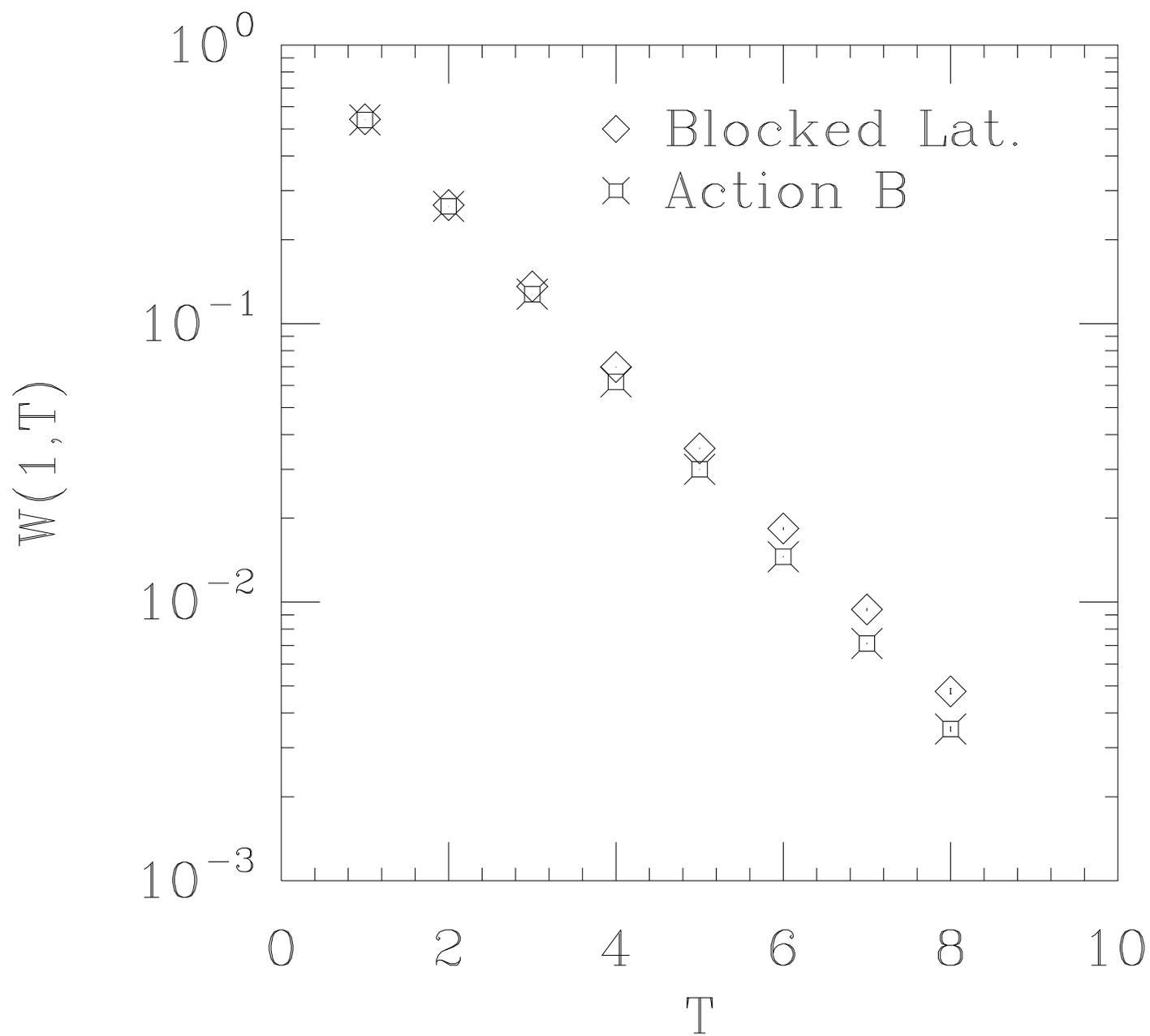}}
\caption{ Same as in Fig.2 but for action {\bf B}.}
\end{figure}

\begin{figure}[htb]
\centerline{\psfig{figure=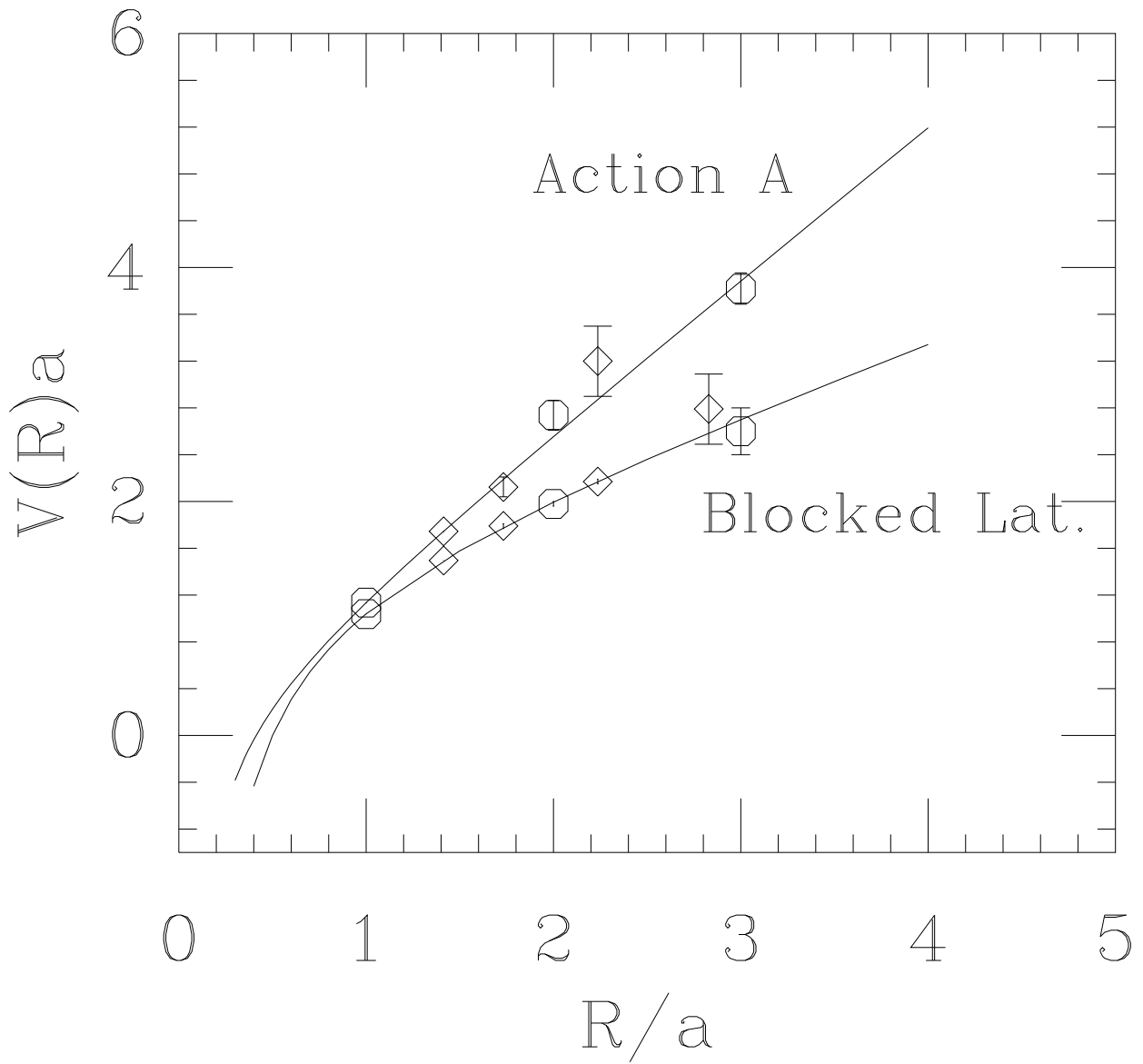}}

\caption{ Heavy quark potential for action {\bf A} and the original lattice blocked at $\beta=6.00$. 
The circles(diamonds) indicate on(off)-axis potentials.
The curve in the figure is obtained by a fit to eq.(7).
}
\end{figure}

\begin{figure}[htb]
\centerline{\psfig{figure=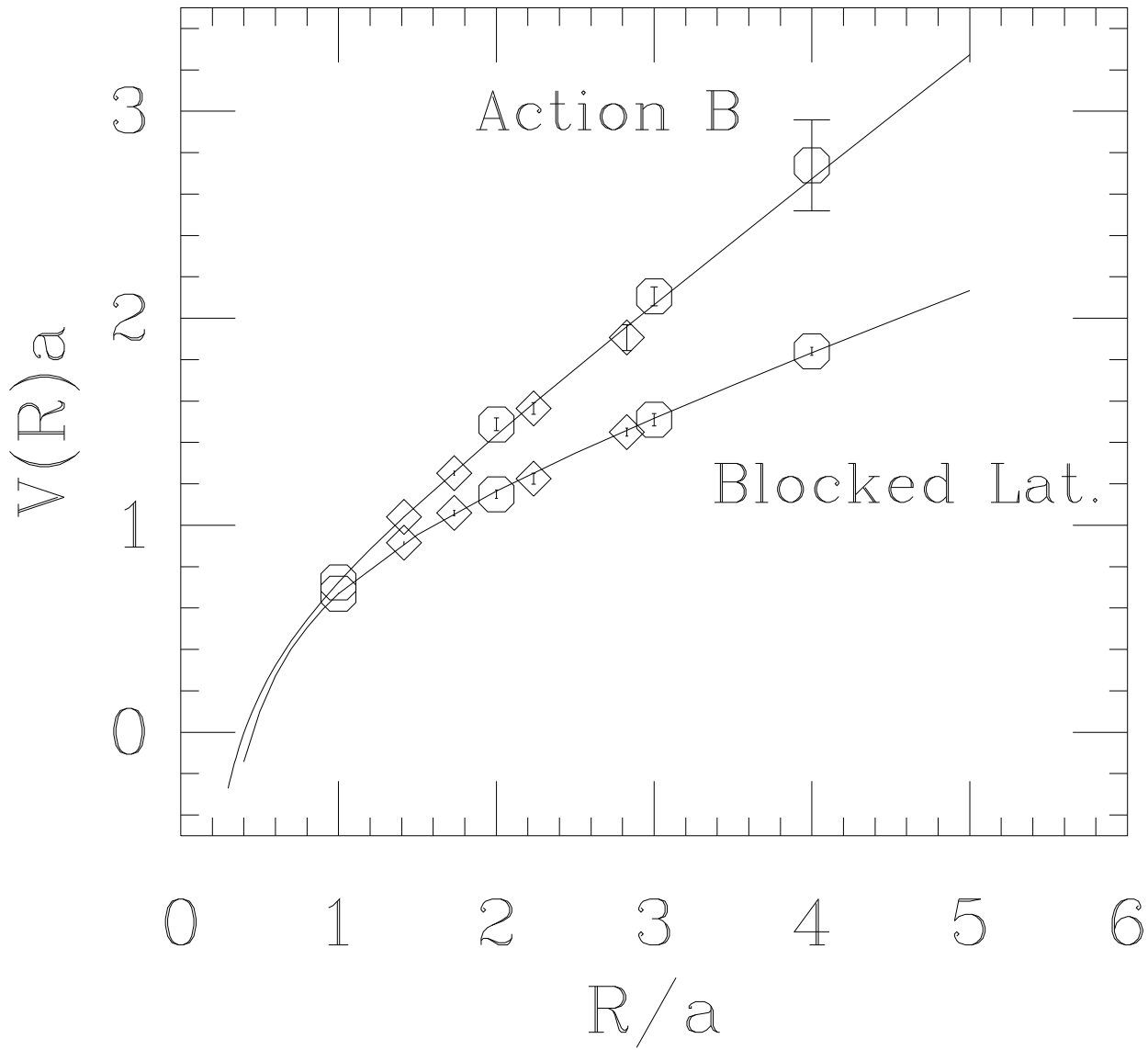}}
\caption{ Same as in Fig4 but for action {\bf B} and the original lattice blocked at $\beta=6.30$.
}
\end{figure}

\begin{figure}[htb]
\centerline{\psfig{figure=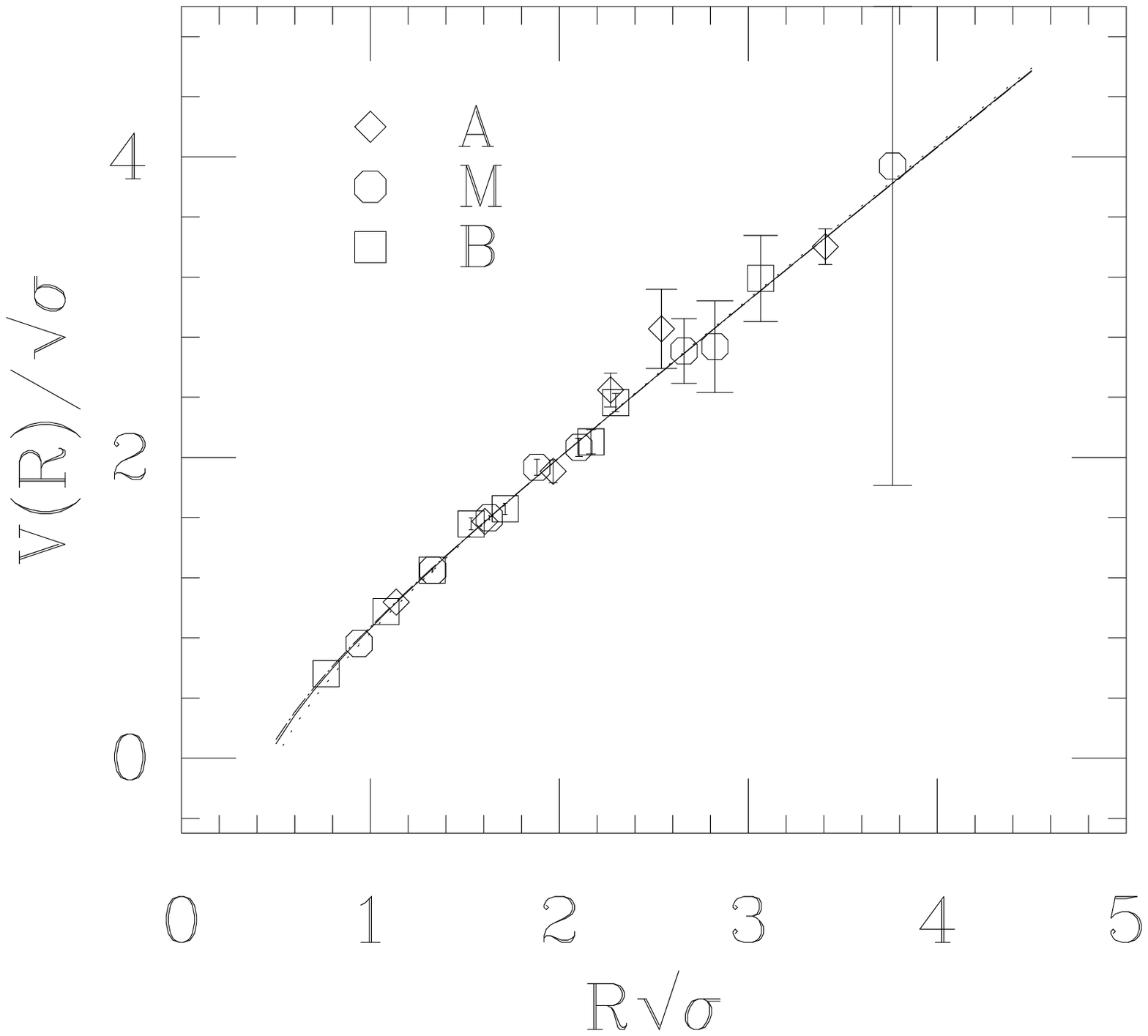}}
\caption{ Heavy quark potentials for actions {\bf A, B} and {\bf M}.
The 3 lines almost on top of each other are separate fits for each action.
}
\end{figure}


\newpage
\begin{thebibliography}{99}
\bibitem{SYM}
K. Symanzik, Nucl. Phys. B226 (1983) 187
\bibitem{WILSON}
K.G. Wilson, Recent progress in gauge theories, Cargese Lectures(1979), eds. G. 't Hooft et al. 
(Plenum, New York, 1980)  
\bibitem{Forcrand}
Ph. de Forcrand and C. Roiesnel, Phys. Lett. B 137 (1984) 213; B 143 (1984) 453
\bibitem{TAD}
G.P Lepage and P.B Mackenzie, Phys. Rev. D 48 (1993) 2250;

M. Alford, W. Dimm, G.P Lepage G. Hockney and P.B Mackenzie, Phys. Lett. B 361 (1995) 87

\bibitem{PERFECT}
P. Hasenfratz and F. Niedermayer, Nucl. Phys. B414 (1995) 785;

T. DeGrand, A Hasenfratz, P Hasenfratz and F. Niedermayer, Nucl. Phys. B454 (1995) 587 ; B454 (1995) 615;
Phys. Lett. B 365 (1996) 233

M. Blatter and F. Niedermayer,  Nucl. Phys. B482 (1996) 286 

\bibitem{ATT}
R.Gupta and A.Patel, Nucl. Phys. B 251 (1985) 789;

R.Gupta et al., Phys. Rev. Lett. 53 (1984) 1721;

M.Creutz, A. Gocksch, M.Ogilvie and M.Okawa, Phys. Rev. Lett. 53 (1984) 875;

P.Stolorz, Phys. Lett. B 172 (1986) 77
\bibitem{MCRG}
R.H. Swendsen, Phys. Rev. Lett. 42 (1979) 859; 47 (1981) 1775

\bibitem{BETA}
See, for example, recent results of \cite{QCDTARO} 
\bibitem{TAKAISHI}
T. Takaishi, Phys. Rev. D 54 (1996) 1050
\bibitem{QCDTARO}
QCD-TARO Collaboration: K.Akemi et al. Phys. Rev. Lett. 71 (1993) 3063; Nucl. Phys. B ( Proc. Suppl.) 34 
(1994) 246
\bibitem{DEMON}
M. Hasenbusch, K. Pinn and C. Wieczerkowski, Phys. Lett B 338 (1994) 308 
\bibitem{DEMONtaka}
T. Takaishi, Mod. Phys. Lett. A 10 (1995) 503
\bibitem{WEISZ}
P. Weisz, Nucl. Phys. B 212 (1983) 1
\bibitem{IWASAKI}
Y. Iwasaki, Report No. UTHEP-118 (1983); S.Itoh, Y.Iwasaki and T.Yoshie, Phys. Rev. D 33 (1986) 1806
\bibitem{BALI}
G.S.Bali and K.Schilling, Phys. Rev. D47 (1993) 661
\bibitem{FIG}
For example see Fig.2 in \cite{TAKAISHI} for comparison.
\bibitem{Iwasaki2}
Y.Iwasaki, K.Kanaya, T.Kaneko and T.Yoshi{\'e},
Nucl. Phys. B (Proc. Suppl.) 53 (1997) 429 
\bibitem{SD}
A. Gonzales-Arroyo and M. Okawa, Phys. Rev. D 35 (1987)
\bibitem{TARO}
QCD-TARO Collaboration: Ph. de Forcrand et al, 
Nucl. Phys. B (Proc. Suppl.) 53 (1997) 938

\end{thebibliography}
\end{document}